\documentclass[aps,prl,twocolumn,floats,showpacs,floatfix,a4paper]{revtex4}
\usepackage{epsfig}
\usepackage{color}
\usepackage{bm}
\usepackage{latexsym}

\begin{document}
\newcommand{\cc}{{\bf\Large C }}
\newcommand{\hide}[1]{}
\newcommand{\tbox}[1]{\mbox{\tiny #1}}
\newcommand{\half}{\mbox{\small $\frac{1}{2}$}}
\newcommand{\sinc}{\mbox{sinc}}
\newcommand{\const}{\mbox{const}}
\newcommand{\trc}{\mbox{trace}}
\newcommand{\intt}{\int\!\!\!\!\int }
\newcommand{\ointt}{\int\!\!\!\!\int\!\!\!\!\!\circ\ }
\newcommand{\eexp}{\mbox{e}^}
\newcommand{\EPS} {\mbox{\LARGE $\epsilon$}}
\newcommand{\ar}{\mathsf r}
\newcommand{\im}{\mbox{Im}}
\newcommand{\re}{\mbox{Re}}
\newcommand{\bmsf}[1]{\bm{\mathsf{#1}}}
\newcommand{\dd}[1]{\:\mbox{d}#1}
\newcommand{\abs}[1]{\left|#1\right|}
\newcommand{\bra}[1]{\left\langle #1\right|}
\newcommand{\ket}[1]{\left|#1\right\rangle }
\newcommand{\mbf}[1]{{\mathbf #1}}
\definecolor{red}{rgb}{1,0.0,0.0}

\title{Bifurcations in Resonance Widths of an Open Bose-Hubbard Dimer}

\author{
Moritz Hiller$^{1,2,3}$, Tsampikos Kottos$^{1,2}$ and Alexander Ossipov$^4$
}
\affiliation{
$^1$Department of Physics, Wesleyan University, Middletown, Connecticut 06459, USA \\
$^2$MPI for Dynamics and Self-Organization, Bunsenstra\ss e 10, D-37073 G\"ottingen, Germany\\
$^3$Fakult\"at f\"ur Physik, Universit\"at G\"ottingen, Friedrich-Hund-Platz 1, D-37077 G\"ottingen, Germany\\
$^4$Instituut-Lorentz, Universiteit Leiden, P.O. Box 9506, 2300 RA Leiden, The Netherlands
}

\begin{abstract}
We investigate the structure of resonance widths of a Bose-Hubbard Dimer with inter-site 
hopping amplitude $k$, which is coupled to continuum at one of the sites with strength 
$\gamma$. Using an effective non-Hermitian Hamiltonian formalism, we show that by varying 
the on-site interaction term $\chi$ the resonances undergo consequent bifurcations. For 
$\Lambda=k/\gamma\geq 0.5$, the bifurcation points follow a scaling law ${\tilde \chi}_m
\equiv \chi_m N/k = f_{\Lambda}(m-0.5/\Lambda)$, where $N$ is the number of bosons. For 
the function $f_{\Lambda}$ two different $\Lambda$ dependences are found around the 
minimum and the maximum bifurcation point.
\end{abstract}
\pacs{05.60.Gg, 03.65.Nk, 03.75.Lm, 03.75.Kk}

\maketitle


The interplay of intrinsic dynamics with coupling to the continuum, radiation field or to any 
other external influence, such as measurement, is an important subject for various branches 
of modern physics: quantum chaos \cite{FKS06}, nanoscale devices \cite{B97}, quantum optics 
\cite{MW95}, theory of decoherence and quantum computing \cite{NC00} up to nuclear \cite{MW69}, 
atomic and molecular physics \cite{atom} are some of the areas, seemingly far remote from each 
other, that boosted the research in open systems. These systems are often described within the 
"effective non-Hermitian Hamiltonian" formalism \cite{MW69}. The eigenvalues of the effective
Hamiltonian are complex ${\cal E}_n=E_n-i\Gamma_n/2$, with the non-zero imaginary part, describing 
the rate with which an eigenstate of the open system (termed resonance state) decays in time 
due to the coupling to the "outside world". 

With the advent of sophisticated techniques for trapping and transporting ultracold atoms, resonance 
states become an issue also in the context of Bose Einstein Condensates (BEC). Examples include
``atom lasers" \cite{BHE98}, optical tweezers \cite{G02} and magnetic ``conveyor belts" \cite{HRHH01}, 
as well as microtraps on ``atom chips" \cite{F00}, which were suggested as potential building 
blocks for quantum information processing \cite{SFC02}. In this context, the aim is to understand 
the scattering and decay properties of a condensate leaking out of the trap via tunneling.

First theoretical investigations of the waveguide scattering of a BEC through a double barrier 
potential, acting like a quantum dot for the atoms, reveal intriguing nonlinear effects associated 
with the internal resonances of the quantum dot \cite{PRS05}. Complementary studies on the decay 
of a condensate from one open trapping potential were reported in \cite{MC04,Moi04,SP04}, while 
a first attempt to study the current relaxation of an ultracold BEC periodically driven with a 
standing wave of laser light was performed in \cite{KW04}. The theoretical approaches used in 
these studies were based on the nonlinear Gross-Pitaevskii equation describing the mean-field 
dynamics of the condensate. 

However, a full quantum many-body study of decaying properties of a BEC is lacking. If the 
hopping amplitude between nearby traps is small compared with the excitation energies in higher 
bands, one can use a Bose-Hubbard Hamiltonian (BHH) to describe the condensate quantum 
mechanically. This is the simplest non-trivial model that describes interacting bosons on a 
lattice, and incorporates the competition between kinetic and interaction energy of the bosonic 
system. In this context, the closed two-site system (dimer) has been analyzed thoroughly and
many interesting phenomena have been found \cite{FPZ00,BES90,KBK03,AGFHCO05,TK88}. The richness of 
the results, provides a motivation to go beyond the closed dimer and consider new scenarios 
where even richer behavior should be observed. In this respect, the open quantum dimer promises 
new exciting opportunities, since the coupling to continuum leads to an interesting interplay 
with the internal dynamics. 


In this paper we address such a scenario and study the resonance linewidths $\Gamma$ 
of a BHH coupled to continuum from one of the two sites. One possible realization of this setup 
is two coupled bosonic traps where tunneling to continuum is imposed at one site. Another 
motivation comes from exciton transfer in molecular aggregates in which guest molecules (traps) 
are introduced interstitially and excitons are captured once they appear inside the sphere of 
influence of the traps \cite{BES95}. Using an effective Hamiltonian formalism, we show the 
appearance of consequent bifurcations in the resonance widths $\Gamma_n$ as we change the 
on-site interaction strength $\chi$. The resonances follow a scaling law 
\begin{equation}
\label{scal1}
{\tilde \Gamma}(k,\gamma,N) \equiv {\Lambda\over N k}\Gamma= \Phi_{\Lambda}(\tilde{\chi}),
\quad {\tilde \chi}=\chi N/k
\end{equation}
where $N$ is the total number of interacting bosons, $k$ the hopping amplitude, $\gamma$ the 
coupling to continuum, and $\Lambda\equiv k/\gamma$. In particular, we reveal that if the 
dimensionless ratio $\Lambda \equiv k/\gamma$ is larger than $\Lambda^*=0.5$, then for $N$ 
interacting bosons there are $N/2$ \cite{note1} bifurcation points $\chi_m,\; m=1,\dots,N/2$ 
which follow a scaling law
\begin{equation}
\label{scaling1}
{\tilde \chi}_m\equiv {\chi_m N\over k}=
f_{\Lambda}\left(\mu \right), \quad \mu\equiv {m-0.5/\Lambda\over N},
\end{equation}
where $f_{\Lambda}$ represent a set of one parameter scaling functions which depend on $\Lambda$. 
Two different $\Lambda-$dependencies are found for the upper and for the lower part of the bifurcation 
spectrum. The former is $\Lambda$-independent and thus universal while for the latter we have 
\begin{equation}
\label{scaling2}
{\tilde \chi}_m/{\tilde \chi}^* = g(\mu),\quad {\tilde \chi}^*=\exp(-1.55/\Lambda)
\end{equation}
where $g(\cdot)$ is a universal ($\Lambda$-independent) scaling function. Our results 
(\ref{scal1},\ref{scaling1},\ref{scaling2}) are confirmed numerically and are supported by 
theoretical arguments.


A usual way to treat the coupling of bound states with continuum is to exclude the continuum 
degrees of freedom from consideration by the introduction of an effective Hamiltonian $H^{\rm eff}$ 
acting within the subspace of bound states and implicitly taking into account their interaction 
with the continuum \cite{MW69}. In the case of a dimer trap, it is natural to assume that 
the  probability to escape in the continuum is proportional to the number of the bosons located 
at the site (say the second one) which is coupled to the continuum \cite{note2}. Following this 
line of argumentation we come out with an $H^{\rm eff}$ given by
\begin{equation}
\label{effect}
\hat{H}^{\rm eff}= \frac{\chi}{2}\sum_{i=1}^{2}{\hat n}_i ({\hat n}_i-1) -
k\sum_{i\neq j}\hat{b}_{i}^{\dagger}\hat{b}_{j}-
i\gamma \hat{b}_2^{\dagger} \hat{b}_2\,,
\end{equation}
where $\chi=4 \pi\hbar^2a_sV_{\tbox{eff}}/m_0$ describes the interaction between two atoms on a single site 
($V_{\tbox{eff}}$ is the effective mode volume of each site, $m_0$ is the atomic mass, and $a_s$ is the
$s$-wave scattering length of atoms which can be 
either positive or negative and can be controlled experimentally). The operators ${\hat n}_i=
\hat{b}_i^{\dagger} \hat{b}_i$ count the number of bosons at site~$i$; the annihilation and 
creation operators $\hat{b}_i$ and $\hat{b}_i^{\dagger}$ obey the canonical commutation 
relations $[\hat{b}_i,\hat{b}_j^{\dagger}] =\delta_{i,j}$. We note that similar effective
Hamiltonian was used to study a leaking dimer \cite{KS02} in the presence of pumping or 
driving light field \cite{SW97}. The Hamiltonian (\ref{effect}) in the fixed (instantaneous) 
basis of the eigenstates of $\hat{N}=\sum_{i=1}^2 \hat{n}_i$ reads
\begin{equation}
\label{efmatrix}
H^{\rm eff}_{nm}=\left\{ \begin{array}{c@{\,:\,}l}
    \frac{\chi}{2}\left[ n^2+(N-n)^2-N\right]
       -i\gamma(N-n) & n=m,\\
    -k\sqrt{n(N+1-n)} & n=m+1,\\
    -k\sqrt{(n+1)(N-n)} & n=m-1,\\
    0 & \mbox{else}.
  \end{array} \right.
\end{equation}
where $n,m=0,1,\dots,N$. We are interested in its (instantaneous) complex eigenvalues ${\cal 
E}_n = E_n - i {\Gamma_n \over 2}$, where $E_n$ and $\Gamma_n$ are the position and the width 
of the resonances, respectively. The subindex $n$ indicates the number of bosons on the first 
(non-dissipative) site. Accordingly $N-n$ bosons are on the second (dissipative) site. The rank 
of $H^{\rm eff}$ is related to the number of particles as ${\cal N}=N+1$. In our numerical 
analysis we consider traps with number of particles $N=10$ up to $500$.

\begin{figure}
\includegraphics[width=\columnwidth,keepaspectratio,clip]{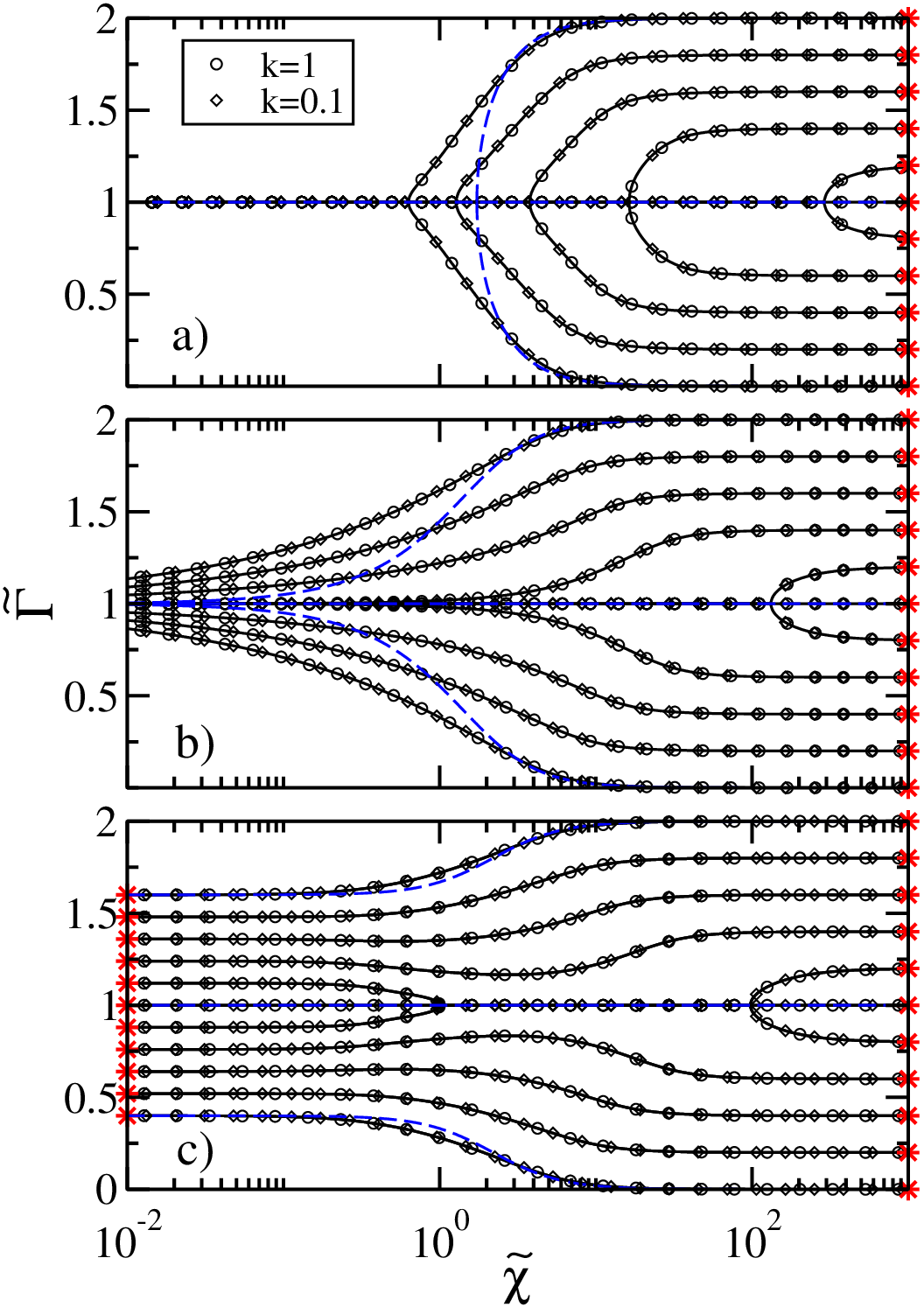}
\caption{\label{fig1}
Bifurcation diagram for the rescaled resonance widths ${\tilde \Gamma}$, vs. the interaction 
strength ${\tilde \chi}$ for three representative coupling ratios: (a) $\Lambda=1$, (b) 
$\Lambda=0.5$ and (c) $\Lambda=0.4$. Circles ($\circ$) and diamonds ($\diamond$) correspond to 
$k=1$ and $0.1$ respectively. In this particular example we consider $N=10$. The solid lines are 
drawn for the eye and show the flow of resonance widths as ${\tilde \chi}$ is changing. The stars 
($\star$) indicate the predicted asymptotic values of the resonance widths (see Eq.~(\ref{Echi0},
\ref{EchiI})). The dashed (blue) line is the "classical" result $\Gamma_{\rm cl}^{\pm}$
}
\end{figure}

To gain some insight in the parametric evolution of resonance widths, let us first consider the two 
limiting cases: vanishing interaction $\chi =0$ and very strong interaction $\chi \gg k$. In the former 
case one finds that the eigenvalues ${\cal E}_n$ of the effective Hamiltonian are given by 
\begin{equation}
\label{Echi0}
{\cal E}_n = N \gamma \left(- {i\over 2} + \left(1-2{n\over N}\right)\sqrt{\Lambda^2-{1\over 4}}\right).
\end{equation}
From this expression one can see that for $\Lambda\geq 1/2$ the resonance widths are all 
degenerate and equal to $N\gamma$ (see Fig.~\ref{fig1}a,b). Instead, for $\Lambda<1/2$ 
the eigenvalues ${\cal E}_n$ are imaginary and we do not have any degeneracies for $\chi=0$ 
(see Fig.~\ref{fig1}c). We have checked that whenever the widths show degeneracy the 
corresponding positions of resonances $E_n$ are not degenerate, exhibiting a ``dual" 
behavior as far as bifurcation points are concerned. 

In the opposite limit of $\chi \gg 1$ one can neglect the inter-site hopping term and thus, 
the Hamiltonian (\ref{effect}) becomes diagonal leading to
\begin{equation}
\label{EchiI}
{\cal E}_n = N\gamma\left(\frac{\chi}{2N\gamma}\left[ n^2+(N-n)^2-N\right] -i\left(1-
{n\over N}\right)\right). 
\end{equation}
Therefore, we obtain $N$ equidistant  resonance widths $\Gamma_n=2\gamma(N-n)$ corresponding
to $n$ particles being trapped on the first site (which resembles the resonance trapping phenomena 
known for the non-interacting particles \cite{Sok92}).
 Naively one could think that the initial degeneracy of the resonance widths for $\Lambda\geq 1/2$ 
can be lifted completely by arbitrary small interaction $\chi$ and the resonance widths 
flow continuously to their limiting values by increasing $\chi$. Below we will see that 
surprisingly enough this is not the case and the degeneracy is reduced each time by 2 at 
$N/2$ discrete points.

In Fig.~\ref{fig1} we report the resonance widths for three representative values of the coupling 
ratio $\Lambda=1,0.5$ and $0.4$, as a function of the interaction strength. Our numerical data
are rescaled according to Eq.~(\ref{scal1}). The points corresponding to the same value of $\Lambda$ 
(but different values of $k$) fall onto the same smooth curve. In the same figure we report the 
asymptotic values given by Eqs.~(\ref{Echi0}, \ref{EchiI}).

In order to shed some light on the emergence of the bifurcations and the scaling ansatz 
(\ref{scal1}) we adopt a semiclassical point of view for Hamiltonian (\ref{effect}), 
justified in the limit of $N\gg 1$. The corresponding classical eigenvalue problem reads
\begin{eqnarray}
\label{nldim}
{\cal E}a_1 &=&\chi \mid a_1\mid^2 a_1 -k a_2, \\
{\cal E}a_2 &=&\chi \mid a_2\mid^2 a_2 -k a_1 -i\gamma a_2,\nonumber
\end{eqnarray}
which can be solved exactly. We find that for $\tilde{\chi}<\tilde{\chi}_{\rm cr}= \sqrt{
4-\Lambda^{-2}}$ there are two ``trivial'' solutions with  constant ($\chi$-independent) 
value of $\Gamma=N\gamma\rightarrow {\tilde \Gamma}=1$ corresponding to the particles 
distributed equally between the two sites $|a_1|^2=|a_2|^2=N/2$. For 
${\tilde \chi}>{\tilde \chi}_{\rm cr}$ two new non-trivial solutions appear with 
$\Gamma_{\rm cl}^{\pm}=\gamma N(1\pm\sqrt{1-4/(\tilde{\chi}^2+\Lambda^{-2})})$ corresponding to 
the non-equal occupations $|a_1^\pm|^2={N\over 2}(1\mp\sqrt{1-4/(\tilde{\chi}^2+\Lambda^{
-2})})$. The classical results are presented also in Fig.~\ref{fig1} (see dashed lines).
This pitchfork bifurcation is the classical analog of the quantum bifurcations discussed 
above. 

\begin{figure}
\includegraphics[width=\columnwidth,keepaspectratio,clip]{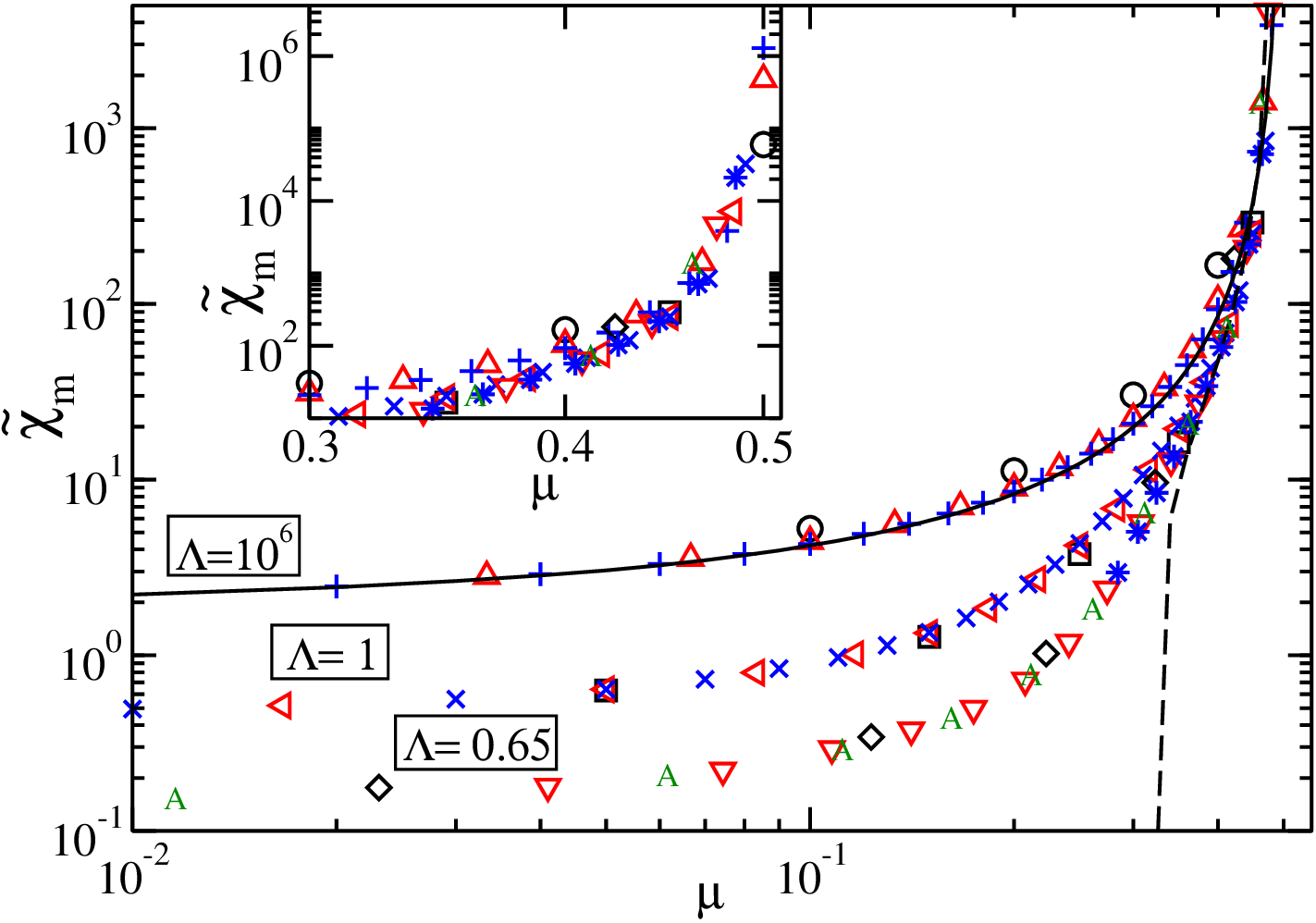}
\caption{\label{fig2}
(Color online) The bifurcation points $\tilde{\chi}_m$ plotted against 
the scaled index $\mu$ for representative values of
the dimensionless ratio $\Lambda=0.65,1,10^6$. The symbols correspond to 
different boson numbers $N$ and
different hopping amplitudes $k$: $N=10,k=1$ (black), $N=30,k=10$ (red), 
$N=50,k=20$ (blue).
The symbol A corresponds to $N=20,k=0.1$.
The solid line indicates the large $N$ limit of the case $\Lambda=10^6$,
while the dashed line corresponds to the value $\Lambda=0.5$ and a 
system size of $N=50$.
In the inset we show a magnification of the main figure.
}
\end{figure}

We see further that for $\Lambda\geq 0.5$ resonances are initially degenerate, as it is 
predicted by Eq.(\ref{Echi0}). As ${\tilde \chi}$ increases this 
degeneracy is lifted by consequent bifurcations. Contrary, for $\Lambda\leq 0.5$ we do not 
observe any initial degeneracy. We note that in the regime $\Lambda < 0.5$ 
the instantaneous approximation is questionable since the hopping time $\tau_k\sim 
1/k$ is large in comparison with the characteristic decay time $\tau_{\gamma}\sim 1/\gamma$. 
Therefore, in the rest of the paper we will concentrate in the opposite case $\Lambda> 0.5$ and analyze 
the sequences of bifurcation points ${\tilde \chi}_m$ which we have extracted from our numerical 
simulations. They were identified as the points after which two degenerate resonance widths 
${\tilde \Gamma}$ differ by more than $C/N$, where we used $C=0.01$. In Fig.~\ref{fig2} we 
report our findings by making use of the scaling proposed in Eq.~(\ref{scaling1}). An excellent 
agreement is evident. We have checked that our results remain qualitatively the same if we 
apply other similar criteria. A side remark is that the resonance positions $E_n$ (not shown) 
bifurcate at the same points ${\tilde \chi}_m$.


Now we focus on the dependence of the scaling function $f_\Lambda$ on the parameter $\Lambda$.
Our study has revealed the existence of a sort of transition separating two regimes characterized 
by different scaling properties. Namely, we found that
at large $\mu$-values (i.e. large ${\tilde \chi}$) the various $\Lambda$-curves fall nicely 
one on top of another, even for $\Lambda=\Lambda^*$, as clearly demonstrated in the inset 
of Fig.~\ref{fig2}. From our data we were able to estimate that this scaling of the upper 
part of the spectrum holds for $\mu \ge 0.35$. Notice that in the limit of $N\rightarrow 
\infty$ at the fixed ratio $m/N$ the scaling parameter $\mu \approx m/N$ i.e. is  independent 
of $\Lambda$. In other words the bifurcations with large $m$ values take place between states 
that are not affected by the inter-site coupling $k$ and the coupling to continuum $\gamma$. 
Indeed these states correspond to almost equidistributed number of bosons between the two traps.

\begin{figure}
\includegraphics[width=\columnwidth,keepaspectratio,clip]{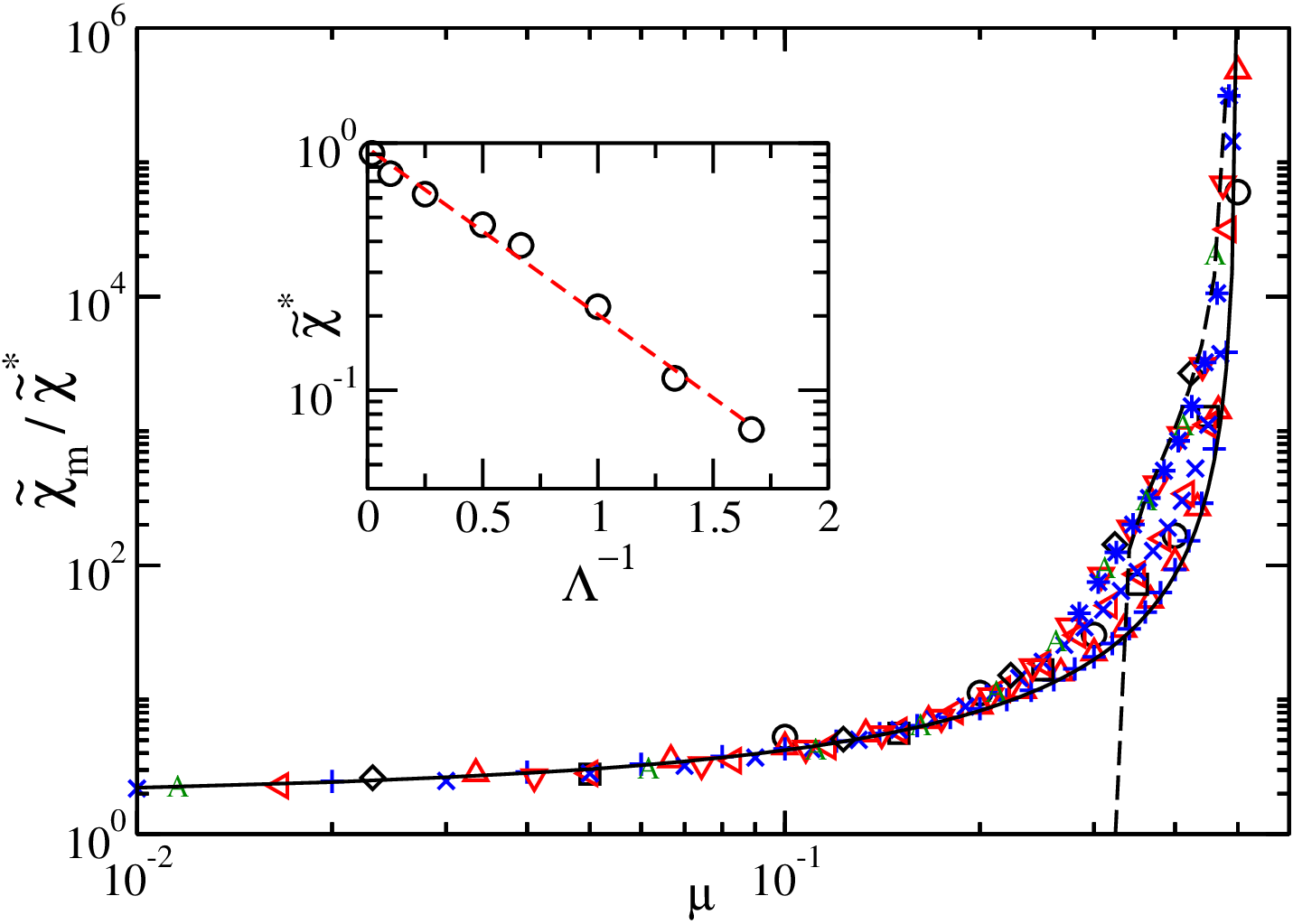}
\caption{\label{fig3}
(Color online) The rescaled bifurcation points 
$\tilde{\chi}_m/\tilde{\chi}^*$ for three values of $\Lambda=0.65,1,10^6$.
The symbols correspond to different boson numbers $N$ and different 
hopping amplitudes $k$ (see Fig.~2).
The solid line indicates the large $N$ limit of the case $\Lambda=10^6$.
{\em Inset}: the extracted scaling parameter $\tilde{\chi}^*$ as a 
function of the dimensionless ratio $\Lambda^{-1}$.
}
\end{figure}

On the other hand, for $\mu\leq 0.3$ we see that all curves, corresponding to different 
values of $\Lambda$ have the same functional form, albeit being shifted downwards with respect to each 
other (Fig.~\ref{fig2}). The curves ${\tilde \chi}$ do coincide, however, when rescaling the 
bifurcation spectra as ${\tilde \chi}_m/{\tilde 
\chi}^*$ where ${\tilde \chi}^*$ is a scaling parameter (Fig.~\ref{fig3}, main part). We 
find that the scaling parameter~${\tilde \chi}^*$ depends only on $\Lambda$ (Fig.~\ref{fig3}, 
inset) resulting in a universal one-parameter scaling for the lower part of bifurcation points 
according to Eq.~(\ref{scaling2}). The best fitting indicates that ${\tilde \chi}^*(\Lambda)
\approx \exp(-1.55\Lambda)$. We note that the scaling region becomes increasingly small as $\Lambda
\rightarrow \Lambda^*$ and breaks down totally at $\Lambda=\Lambda^*$. This is the limit where 
the resonance widths do not show any degeneracy for $\chi=0$.

In conclusion we have studied the dependence of resonance widths of an open dimer BHH system 
on the strength of the on-site interaction and found a rich bifurcation behavior. Our results
have immediate implications in the dynamics of an open dimer. Specifically, we expect that
the decay of the bosons is not given by a simple exponential as suggested by the standard rate 
equation $\dot{N}=-\Gamma N$ with $\Gamma\sim \gamma$ being a constant. Instead, our analysis 
indicate a more complicated decay since now $\Gamma$ depends on the remaining particles inside 
the trap (see Eq.(\ref{scal1}) and Fig.~1) and thus one has to solve a non-linear rate equation. 
The analysis of dynamics of an open BHH system is beyond the scope of the current publication 
and will be the subject of a separate paper.

\begin{thebibliography}{99}
\bibitem{note1} For odd numbers of bosons, there are $(N+1)/2$ bifurcation points.
\bibitem{note2} We can derive Eq. (\ref{effect}) starting from the classical 
Hamiltonian ${\cal H}=\frac{\chi}{2}\left [ \mid a_1\mid^4+\mid a_2 \mid^4\right]- 
k(a_1^*a_2+a_1a_2^*)-i\gamma a_2^*a_2$, where $a_i$ are complex amplitudes, 
and then apply second 
quantization. Here, constant terms due to symmetrization are neglected.
For $\gamma=0$ this leads to the standard Discrete Nonlinear Schr\"odinger 
(DNLS) equation while for $\gamma\neq 0$ it leads to the DNLS with a sink
 \cite{BES95}.
\end{thebibliography}
\end{document}